\documentclass{article}
\usepackage{spconf,amsmath,graphicx}
\usepackage{comment, array, bm, pifont, mathtools, float, graphicx, xcolor, booktabs, xfrac, scalerel, enumerate, enumitem, multirow, pdfrender, placeins, support-caption}
\usepackage[caption=false]{subfig}
\usepackage{tikz, pgfplots}
\usepackage[export]{adjustbox}
\usepackage{setspace,dashbox}
\usepackage{hyperref}
\definecolor{brown(web)}{rgb}{0.65, 0.16, 0.16}
\definecolor{brightmaroon}{rgb}{0.76, 0.13, 0.28}
\definecolor{applegreen}{rgb}{0.03, 0.47, 0.19}
\hypersetup{
colorlinks=True,
citecolor=brown(web),
urlcolor=blue,
linkcolor=brown(web)}
\usepackage[numbers,sort]{natbib}
\usepackage[a-1b]{pdfx} 

\newcommand\dashedph[1][H]{\setlength{\fboxsep}{0pt}\setlength{\dashlength}{2.2pt}\setlength{\dashdash}{1.1pt} \dbox{\phantom{#1}}}

\pgfplotsset{compat=1.16} 
\DeclarePairedDelimiter\autobracket{(}{)}
\newcommand{\br}[1]{\autobracket*{#1}}

\title{Deep Quantized Representation for Enhanced Reconstruction}
\name{Akash Gupta$^{\dagger, \star}$, Abhishek Aich$^\dagger$, Kevin Rodriguez$^\ddagger$, G. Venugopala Reddy$^\ddagger$, Amit K. Roy-Chowdhury$^\dagger$\thanks{This work was partially supported by NSF grants 1664172 from the Office of Advanced Cyberinfrastructure and 1762063 from the Division of Mathematical Sciences.~$^\star$indicates corresponding author. E-mails: \texttt{\{agupt013@, aaich001@, krodr005@, venug@, amitrc@ee.\}ucr.edu}.}}
\address{
$^\dagger$Department of Electrical and Computer Engineering, $^\ddagger$Department of Botany and Plant Sciences,\\ University of California, Riverside}

\begin{document}

\maketitle
\begin{abstract}
While machine learning approaches have shown remarkable performance in biomedical image analysis, most of these methods rely on high-quality and accurate imaging data. However, collecting such data requires intensive and careful manual effort.
One of the major challenges in imaging the Shoot Apical Meristem (SAM) of Arabidopsis thaliana, is that the deeper slices in the $z-$stack suffer from different perpetual quality related problems like poor contrast and blurring. 
These quality related issues often lead to disposal of the painstakingly collected data with little to no control on quality while collecting the data. Therefore, it becomes necessary to employ and design techniques that can enhance the images to make it more suitable for further analysis. 
In this paper, we propose a data-driven Deep Quantized Latent Representation (DQLR) methodology for high-quality image reconstruction in the Shoot Apical Meristem (SAM) of Arabidopsis thaliana. Our proposed framework utilizes multiple consecutive slices in the $z$-stack to learn a low dimensional latent space, quantize it and subsequently perform reconstruction using the quantized representation to obtain sharper images. Experiments on a publicly available dataset validate our methodology showing promising results. Our code is available at  \href{https://github.com/agupt013/enhancedRec.git}{\texttt{github.com/agupt013/enhancedRec.git}}.
\end{abstract}
\vspace{-0.7em}
\section{Introduction}
\label{sec:intro}

\begin{figure}[!t] 
    \centering
    \includegraphics[width=0.47\textwidth, height=2.6cm]{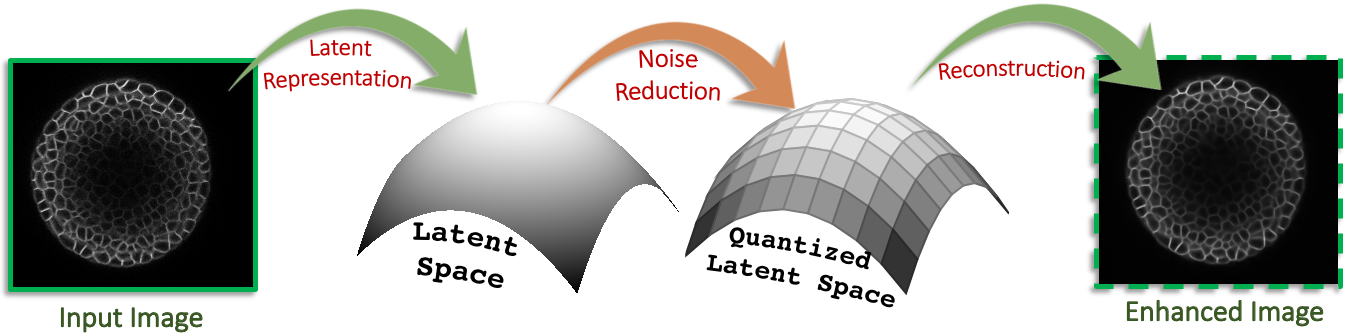}
    \vspace{-1em}
    \caption{\textbf{Conceptual Overview of DQLR.} The latent representation of the collected image is quantized using $k-$means over the entire dataset~\cite{van2017neural}. This quantized representation is then used to reconstruct the enhanced image.}
    \label{fig:concept}
    \vspace{-2em}
\end{figure}

Automated analysis in biomedical research is critical to provide researchers with concrete evidence to prove any proposed hypothesis without any bias. However, automated image analysis requires high-quality imaging data. Image quality related problems are often encountered while imaging deeper layers of the Shoot Apical Meristem (SAM) of arabidopsis thaliana ~\cite{liu2011adaptive}.  These quality related problems hinder automated analysis and often lead to disposal of painstakingly collected data. To this end, we propose a data driven Deep Quantized Latent Representation (DQLR) framework for high-quality imaging data reconstruction of the $z-$stack of the SAM. In this work, we propose to project noisy stack in a latent space, quantize the latent representations and utilize the quantized latent representations for reconstruction of enhanced $z-$stack (see Fig.~\ref{fig:concept} for conceptual overview). 

\textbf{Overview.} An architectural overview of our approach is illustrated in Fig.~\ref{fig:overview}. During training, the encoder $\mathsf{E}$ compresses $i^{th}$ input slice image to a latent representation $x_i$. 
The consecutive slices in the $z$-stack are correlated which implies that they must be correlated in the latent space as well. We employ a recurrent neural network (RNN) $\mathsf{R}$ to learn this correlated representation $\{y_{i}, y_{i+1}, \cdots, y_{i+n}\}$ by passing the latent vector $\{x_1, x_2, \cdots, x_n\}$ through $\mathsf{R}$.
The compressed representation $x_i$ is processed through $\mathsf{R}_i$ to learn the inter-correlation between this latent representation of the consecutive slices $\{{x_i, x_{i+1}, \cdots, x_{i+n}}\}$ during training. RNN generated latent codes $\{y_{i}, y_{i+1}, \cdots, y_{i+n}\}$ are then used as input to quantization module $\mathsf{Q}_i$.
$\mathsf{Q}_i$ learns a vector dictionary for quantized representation of the network and generates a quantized latent code $\{y^{q}_{i}, y^{q}_{i+1}, \cdots, y^{q}_{i+n}\}$. In our proposed method, the quantization of the latent code will remove the noisy component of $\{y_i\}$ and the reconstructed/predicted images using the quantized latent codes by generator $\mathsf{G}$ should be enhanced. 
During testing, we pass one slice at a time from the $z-$stack, compress it using the encoder, predict the correlated latent codes using the RNN, and finally quantize it using the quantization dictionary learned during the training stage using $\mathsf{Q}$. This quantized code is then used to reconstruct and predict enhanced consecutive slices from the given $z-$stack.
\vspace{-0.7em}
\section{Related Work}
\label{sec:related_work}
In this section we describe prior works closely related to the our proposed method. Our method closely relates to reconstruction using auto-encoders~\cite{schmidhuber2015deep} and enhancement in the compressed domain~\cite{van2017neural, tang2003image, wu2019deep}.\\[-0.6em]

\noindent \textbf{Auto-Encoders.} Variations of auto-encoders are extensively used in reconstruction tasks by compressing the input to a latent representation and  using the latent representation to retrieve the input as close as possible~\cite{van2017neural, schmidhuber2015deep, makhzani2015adversarial}. However, often the reconstructed images are blurry due to inherent nature of Mean Square Error (MSE) loss to produce blurry results. In the proposed approach we also include Structural Similarity Index (SSIM)~\cite{zhao2016loss} 
loss to enhance the visual results.\\[-0.6em]

\noindent \textbf{Compressed Domain Enhancement.} Some works have tried to enhance the images in the compressed domain. In~\cite{tang2003image} a method based on a contrast measure defined within the discrete cosine transform (DCT) domain is proposed to enhance the image. Attention based video enhancement is proposed in~\cite{gupta2020alanet}. Authors in ~\cite{van2017neural} propose a vector quantized variations auto-encoder for reconstruction of various media input. We adopt their approach of vector quantization in our framework. However, we exploit the input data correlation using RNN for enhancement task as opposed to reconstruction in ~\cite{van2017neural} where ground truth data was available.

\section{Methodology}
\label{sec:proposed_method}
\begin{figure*}[t]
    \centering
    \includegraphics[width=0.8\textwidth, height=7.8cm]{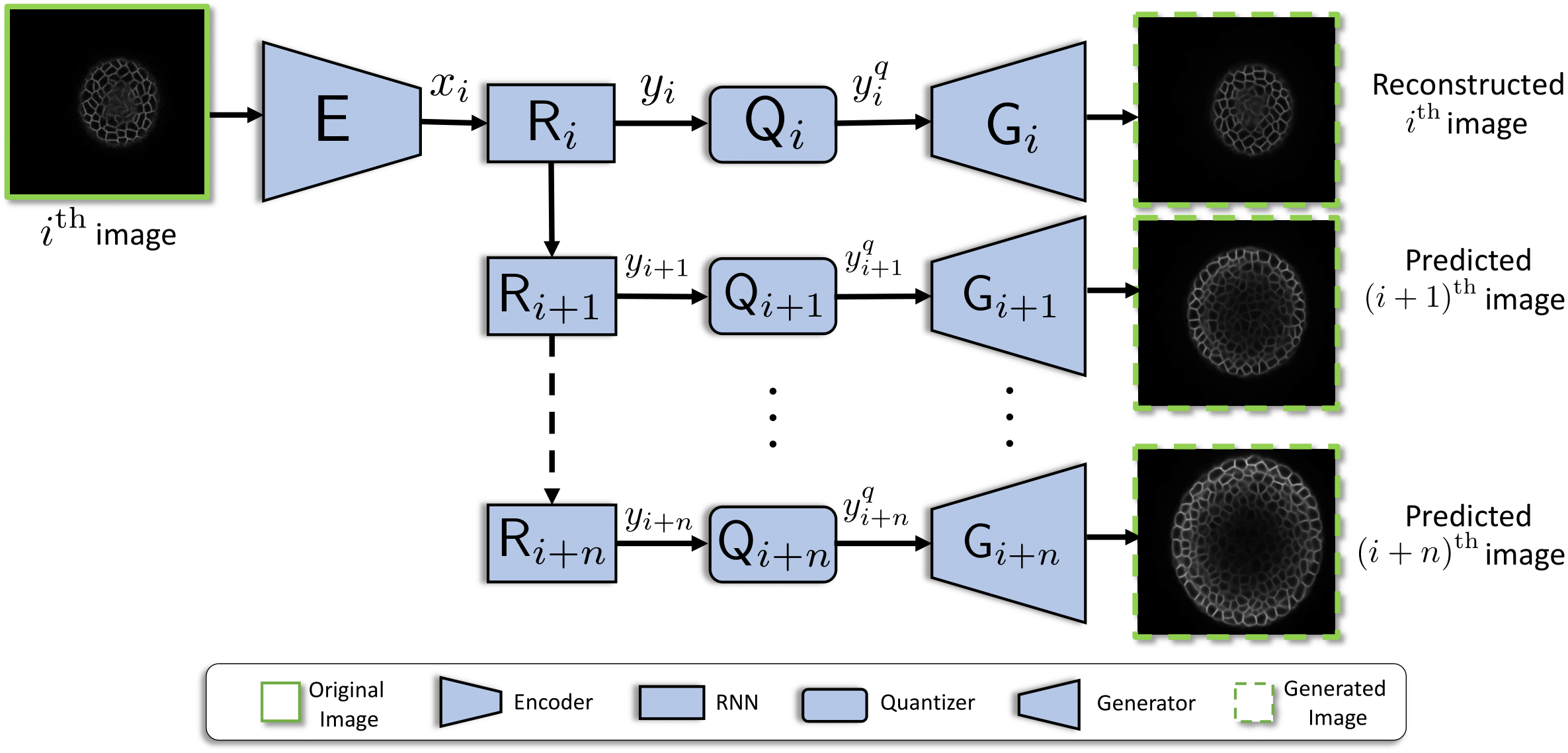}
    \caption{\textbf{Architectural Overview of DQLR (for one slice of the stack).}
    Encoder $\mathsf{E}$ encodes input image to $x_i$. Recurrent Neural Network (RNN) module generates correlated codes for reconstruction ($y_i$) and prediction $\br{\{y_{i}, y_{i+1}, \cdots, y_{i+n}\}}$. Quantizer module $\mathsf{Q}_i$ quantizes the latent codes and Generator $\mathsf{G}$ reconstructs/predicts the images.
    }
    \label{fig:overview}
    \vspace{-1em}
\end{figure*}
We propose a Deep Quantized Latent Representation (DQLR) framework for enhancing $z-$stack imaging in SAM of Arabidopsis thaliana. We apply quantization in the latent space of the noisy $z-$stack for enhanced reconstruction. In this section, we first formulate the problem statement and then explain our proposed approach in details.

\subsection{Problem Formulation}\label{ssec:problem}
Given a $z-$stack $\mathbf{Z}$ = $\{\mathbf{z}_1, \mathbf{z}_2, \cdots, \mathbf{z}_n\}$, with $\textbf{z}_i$ being the $i^{th}$ slice in the stack from the top, we aim to reconstruct $\widehat{\mathbf{Z}}$ = $\{\widehat{\mathbf{z}}_1, \widehat{\mathbf{z}}_2, \cdots, \widehat{\mathbf{z}}_n\}$ such that $\widehat{\mathbf{z}}_i$ is the visually enhanced slice compared to $\mathbf{z}_i, \forall i=1, 2, \cdots, n$. 
Let there be a latent representation of input noisy $z-$stack $\textbf{X}_{\mathbf{Z}} =  \{ x_1, x_2, \cdots, x_n \}$ where $x_i$ is the latent representation corresponding to the $i^{th}$ slice $\mathbf{z}_i$. Since the slices in $z-$stack are correlated in the pixel space, their latent representations should inherit the same property in the latent space. Therefore, corresponding to each latent representation $\textbf{X}_{\mathbf{Z}}$ let there be a latent representation $\textbf{Y}_{\mathbf{Z}} =$  $\{ y_1, y_2, \cdots, y_n \}$ such that all $\{ y_i\}$ are correlated.

We propose to generate visually enhanced $z-$stack by quantizing the latent representation of the noisy input stack. Our hypothesis is that each correlated latent representation $y_{i}$ of a slice in the $z-$stack consists of two components; the quantized representation $y^q_{i}$ and the noise representation $y^\text{noise}_{i}$ of $y_{i}$, such that $y_i = y^q_{i} + y^{\text{noise}}_{i}$. Hence, noise component $y^{\text{noise}}_{i}$ can be removed by applying quantization on the correlated latent codes leaving the representation $y^q_i$ required to generate the enhanced image $\widehat{\textbf{z}}_i ~\forall~i = 1, 2,\cdots, n$.

%
\subsection{Proposed Approach}\label{ssec:dqlr}
Our proposed framework is shown in Figure~\ref{fig:overview}. It consists of four components: the encoder network $\mathsf{E}$, the recurrent neural network $\mathsf{R}$, the quantization module $\mathsf{Q}$ and the generator network $\mathsf{G}$. The encoder network is used to extract latent representation for each slice in the noisy input stack. The recurrent neural network utilizes the latent representations to generate correlated latent representations. These correlated representations are quantized to reduce noise in the latent space by the quantization module. Finally, the quantized representations are used to generate an enhanced $z-$stack.\\[-0.8em]

\noindent \textbf{Input Latent Representation.} We employ a convolutional neural network as an encoder $\mathsf{E}$ which extracts the latent representation for each slice in a given noisy $z-$stack such that
\begin{align}
    \mathsf{E}\br{\mathbf{Z}}
    = \mathsf{E}\br{ \{ ~\textbf{z}_1, ~\textbf{z}_2, \cdots, ~\textbf{z}_n \}}
    =\{ ~x_0, ~x_1, \cdots, ~x_n \} 
\end{align}
where $x_i$ is latent representation corresponding to slice $\textbf{z}_i$. A set of correlated representations is generated by the recurrent neural network for the latent representations extracted from the encoder $\mathsf{E}$ to incorporate the z-resolution dynamics of the $z-$stack in the latent representations.\\[-0.6em]

\noindent \textbf{Recurrent Neural Network (RNN).} The consecutive slices in a $z-$stack capture 3D-structure of any cell in the plant. Thus, there must be a correlation between the consecutive slices. The latent representation $\textbf{X}_\mathbf{Z}$ of the noisy input $\mathbf{Z}$ should also be correlated in some space $\textbf{Y}_\mathbf{Z}$. Thus, we employ a recurrent neural network $\mathsf{R}_i$ to transform the $i^{th}$ noisy latent representation to the correlated latent representation as RNN can capture dynamics of the sequence given by
\begin{align}
    y_{i+1} = \mathsf{R}_i\br{y_i, h_i}
\end{align}
\noindent where $h_0$ is the hidden state sampled randomly from a Gaussian distribution and $h_i = x_{i-1} ~\forall ~i > 0$. Here, we aim to capture the $z-$resolution dynamics of the stack unlike traditional recurrent neural network where temporal dynamics of the sequence is captured.\\

\noindent \textbf{Deep Quantized Latent Representation.} We propose that a data driven quantization of the latent representation can reduce the average noise in the stack and enhance it visually. In order to quantize the latent representation, we employ vector quantization dictionary learning algorithm as proposed in ~\cite{van2017neural}, represented as $\mathsf{Q_i}$ in our framework.\\

\noindent \textbf{Enhanced Stack Generation.}
We employ a generative model $\mathsf{G}$ to transform the quantized representations into an enhanced stack $\widehat{\mathbf{Z}}$. The quantized representations $\textbf{Y}^q_{\mathbf{Z}}$ are used by the generator $\mathsf{G}$ to synthesize enhanced stack $\widehat{\mathbf{Z}} = \{\widehat{\mathbf{z}}_1, \widehat{\mathbf{z}}_2, \cdots, \widehat{\mathbf{z}}_n\}$ such that $\widehat{\mathbf{z}_i}$ is the visually enhanced image of the slice $\mathbf{z}_i$ in the noisy stack $\mathbf{Z}$.

\subsection{Optimization}
Our optimization function consists of the Mean Squared Error (MSE) pixel reconstruction loss, the Structural Similarity (SSIM) loss~\cite{zhao2016loss} and quantization loss as defined in~\cite{van2017neural}. Please note that we do not have de-noised image as ground truth. We assume that the quantized latent codes should reduce noise when it is used by generator $\mathsf{G}$ to reconstruct the stack. Results in section~\ref{sec:experiment_results} demonstrate the validity of this assumption. 
\begin{align}\label{eqn:gan}
    \mathcal{L}_{\text{total}} = \mathcal{L}_{\text{mse}} +  \lambda_{\text{s}}\mathcal{L}_{\text{ssim}} + \lambda_{\text{q}} \mathcal{L}_{\text{quant}}
\end{align}

We briefly describe the loss functions below. Define $P$ as the total number of non-overlapping patches in a given image, $N$ as total number of pixels in $P$, and $\alpha$ and $\beta$ as the generated and ground truth image, respectively.  
\begin{gather*}
    \mathcal{L}_{\text{mse}}(P) = \dfrac{1}{N}\sum_{p\in P} \Vert \alpha(p) - \beta(p)\Vert_2 \\
    \mathcal{L}_{\text{ssim}}(P) = \dfrac{1}{N}\sum_{p\in P} 1 - \text{SSIM}(p),\\
    \text{with},~\text{SSIM}(p) = \br{\dfrac{2\mu_\alpha\mu_\beta + C_1}{\mu_\alpha^2 +\mu_\beta^2 + C_1}}\br{\dfrac{2\sigma_\alpha\sigma_\beta + C_2}{\sigma_\alpha^2 +\sigma_\beta^2 + C_2}} 
\end{gather*}
where,  $\mu_{\br{\cdot}}$ and $\sigma_{\br{\cdot}}$ are computed with a Gaussian filter with standard deviation $\sigma_G$, $C_1 < 1$ and  $C_2 < 1$ are constants introduced to handle division by zero issue, $\lambda_{\text{s}}$ and $\lambda_{\text{q}}$ weights for SSIM and quantization loss, respectively. For $\mathcal{L}_{\text{quant}}$, we use the loss function as proposed in~\cite{van2017neural} on the correlated latent space $\textbf{Y}_\mathbf{Z}$ and dictionary $\mathbf{D} = \{ d^1, d^2,\cdots, d^k \}$, where $k = 128$ is length of dictionary to learn for quantization.
\section{Experimentation and Results}
\label{sec:experiment_results}
\noindent \textbf{Datasets.} We used the publicly available Confocal Membrane dataset~\cite{willis2016cell} consisting of six plants. We train our model using four plant stacks, and use one plant stack each for validation and testing.\\

\begin{figure}[!htp]
    \vspace{-0.5em}
    \centering
    \subfloat{
    \includegraphics[width = \columnwidth]{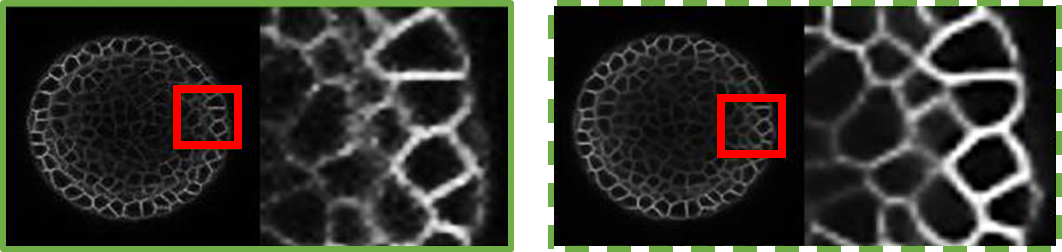}
    }
    \vfill
    \subfloat{
    \includegraphics[width = \columnwidth]{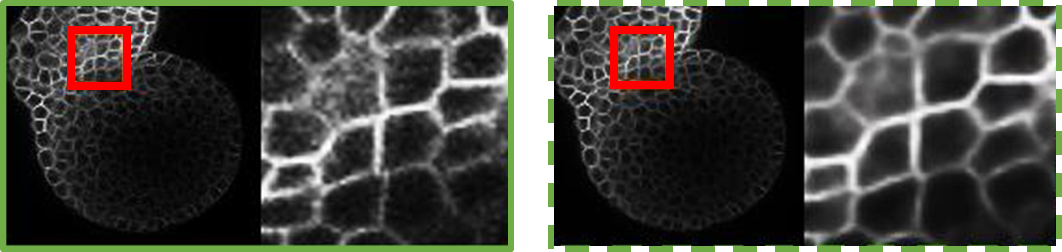}
    }
    \vspace{-0.5em}
    \caption{\textbf{Qualitative Results of Proposed Method.} Original image (\textit{left}) and Reconstructed image (\textit{right}) with corresponding zoomed parts are presented here. The proposed method is able to generate sharper images from the given blurry image slices.}
    \label{fig:main}
    
\end{figure}

\begin{figure}[H]
    \vspace{-1em}
    \centering
    \includegraphics[width = \columnwidth]{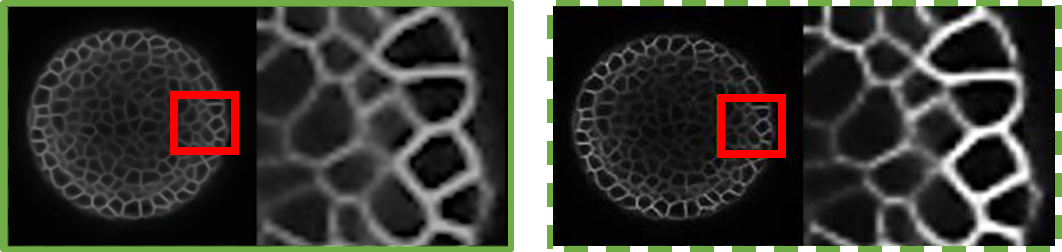}
    \vspace{-1.5em}
    \caption{\textbf{Reconstruction Results without Quantization.} Reconstructed image without quantization (\textit{left}) and Reconstructed image (\textit{right}) with quantization with corresponding zoomed parts are presented here. This demonstrates that the quantization module in our proposed approach is effective in deblurring the data.}
    \label{fig:no_quant}
    
\end{figure}

\begin{figure}[H]
    \vspace{-1em}
    \centering
    \includegraphics[width = 0.97\columnwidth]{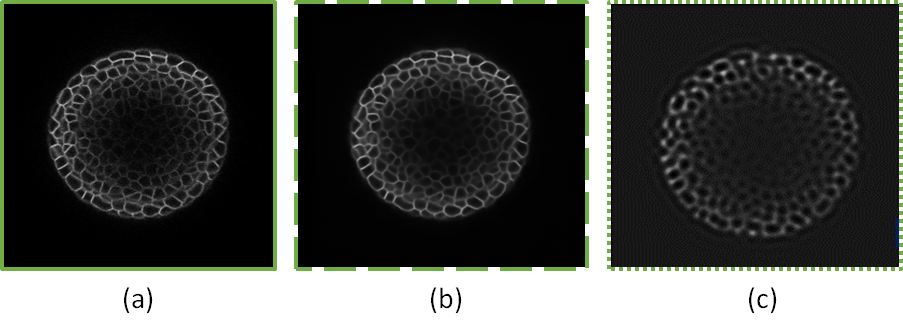}
    \vspace{-1.5em}
    \caption{\textbf{Comparison of Reconstructed Results with ImageJ}~\cite{collins2007imagej}.\textbf{(a)} Original Image, \textbf{(b)} Reconstructed using DQLR (\textbf{ours}) and \textbf{(c)} Reconstructed using deconvolution by ImageJ.}
    \label{fig:deconv_compare}
    
\end{figure}

\noindent \textbf{Qualitative Results.} Fig.~\ref{fig:main} shows few examples of the reconstructed slices from the $z-$stack using the our approach along with the input slice. It can be observed that our proposed method is able to generate sharper cell boundaries. Since we learn the quantization dictionary using all the slices in various $z-$stacks, our method is able to generate cleaner images. Deconvolution is a standard technique used by many researchers to enhance  microscopy images. We compare our proposed approach with deconvolution operation used to denoise microscopy images using ImageJ~\cite{collins2007imagej}. It is performed on 2D slices using Gaussian Point Spread Function (PSF) with standard values. It can be seen from Fig.~\ref{fig:deconv_compare} that our proposed approach reconstructs visually enhanced slices compared to deconvolution operation in ImageJ. A key reason that deconvolution doesn't work well is due to the selection of PSF which highly depends on the capturing instrument. This demonstrates the advantage of our approach with respect to existing algorithms. Note that in Fig.~\ref{fig:main}, Fig.~\ref{fig:no_quant}, and Fig.~ \ref{fig:deconv_compare}, input slice is shown inside \raisebox{0.9ex}{\resizebox{!}{1ex}{\fcolorbox{applegreen}{white}{\null}}} and the reconstructed slice using the proposed approach is shown inside \raisebox{0.0ex}{\resizebox{!}{2ex}{\textcolor{applegreen}{$\textbf{\dashedph}$}}}. Results are best viewed when zoomed-in.

\noindent \textbf{Qualitative Ablation.} To evaluate the impact of quantization in the latent space, we perform an experiment without applying quantization keeping all other parameters same in the proposed method.  Fig.~\ref{fig:no_quant} qualitatively shows the contribution on quantization in latent space. The image generated without quantization is less sharp than with quantization. This is due to inherent property of mean square loss to produce blurry results which dominates the reconstruction in absence of latent representation quantization loss.

\section{Conclusion}
\label{sec:conclusion}
Micro-imaging data collected for various bio-medical research suffers from inherent blurriness and using this data for further analysis is a challenging task. We present an approach for enhanced reconstruction of microscopic sequential data by leveraging the information from consecutive image slices and using quantization of their latent representation to alleviate blurriness. Our data driven approach demonstrates visually superior results on a publicly available benchmark. The proposed approach would be useful for bio-medical researchers to enhance images where data is scarce and consequently, avoid unwanted laborious efforts for re-imaging the data.\\

\small{
\noindent \textbf{Acknowledgement.} We thank Prof. B.S. Manjunath from University of California, Santa Barbara for valuable discussions and helpful suggestions.}


\bibliographystyle{IEEEbib}
\bibliography{refs}

\end{document}